\documentclass[aps,prd,reprint,twocolumn,preprintnumbers,floatfix,nofootinbib]{revtex4-1}
\usepackage[utf8]{inputenc} 
\usepackage[dvips]{graphicx}
\usepackage[dvipsnames]{xcolor}
\usepackage{color}
\usepackage{relsize}
\usepackage{float}
\usepackage{graphics}
\usepackage{epstopdf}
\usepackage{hyperref}
\usepackage{mathrsfs}
\usepackage{amsfonts}
\usepackage{dcolumn}
\usepackage{bm}
\usepackage{physics}
\usepackage{booktabs}
\renewcommand{\figurename}{Fig.}
\usepackage[normalem]{ ulem }
\usepackage{amsthm}
\usepackage{amssymb}
\usepackage{amsmath}
\usepackage{cancel}
\usepackage{fontawesome} 
\usepackage[caption=false]{subfig}
\usepackage{tabularx,ragged2e} 
\usepackage{lipsum}
\newcolumntype{L}{>{\RaggedRight\arraybackslash}X}

\usepackage{hyperref}

\newcommand{\figref}[1]{Fig.~\ref{#1}}

\newcommand{\refref}[1]{Ref.~\cite{#1}}

\allowdisplaybreaks[4]

\newcommand\myshade{80}
\colorlet{mylinkcolor}{ForestGreen}
\colorlet{mycitecolor}{Red}
\colorlet{myurlcolor}{violet}

\hypersetup{
	linkcolor  = mylinkcolor!\myshade!black,
	citecolor  = mycitecolor!\myshade!black,
	urlcolor   = myurlcolor!\myshade!black,
	colorlinks = true
}

\newcommand{\calA}{\mathcal{A}}
\newcommand{\calAS}{\mathcal{A}(S,\overline{S})}
\newcommand{\mathfa}{\mathfrak{a}}

\newcommand{\I}{{\rm i}}

\graphicspath{{Figures/}}

\begin{document}

\title{Modular Invariant Hilltop Inflation}

\author{Stephen F. King}
\email{King@soton.ac.uk}
\author{Xin Wang}
\email{Xin.Wang@soton.ac.uk}
\affiliation{School of Physics and Astronomy, University of Southampton, Southampton SO17 1BJ, United Kingdom}

\begin{abstract}
In this paper we show that it is possible to achieve successful hilltop inflation in which the inflaton is identified as the modulus field in a modular invariant theory.
The dilaton plays a crucial role in shaping the potential. 
Modular invariant gaugino condensation provides the mechanism for the modulus stabilisation after inflation. The inflationary trajectory lies on the lower boundary of the fundamental domain of the modulus field $\tau$. Inflation starts near the fixed point $\tau =\I$, and ends at a point near $\tau = \omega$, which is the global de Sitter vacuum.
We investigate the allowed parameter space for successful modular invariant hilltop inflation.
\end{abstract}

\maketitle

\section{Introduction}
Understanding the evolution of the early universe remains a big challenge. Regarding this, inflation~\cite{Brout:1977ix,Starobinsky:1980te,Kazanas:1980tx,Sato:1981qmu,Guth:1980zm,Linde:1981mu,Albrecht:1982wi,Linde:1983gd} provides us with an elegant framework to remedy several defects in the hot big-bang cosmology, e.g., the horizon, flatness, entropy and primordial perturbation problems. The basic idea of inflation is the hypothesis that the early universe underwent a phase of accelerated superluminal expansion, which can be easily achieved in the slow-roll inflationary paradigm~\cite{Linde:1990flp,Lyth:1998xn}. Despite lacking a direct method to detect inflation, there are still some observables relevant to inflation from the measurements of the cosmic microwave background, particularly the tensor-to-scalar ratio $r$ and the spectral index $n_s$. The joint constraints from the Planck 2018 observation and the BICEP/Keck 2018 observing season~\cite{Planck:2018jri,BICEP:2021xfz} indicate that $n_s = 0.9661 \pm 0.0040$ at 68\% CL and $r < 0.036$ at 95\% CL. Among various inflation scenarios, hilltop inflation~\cite{Boubekeur:2005zm, Linde:1981mu, Albrecht:1982wi,Izawa:1996dv,Senoguz:2004ky} is a class of simple models described by the potential $V(\phi) = \Lambda^4 ( 1 - \phi^{p}/\mu^{p} + \cdots )$ of a real field $\phi$ (with $\Lambda$ and $\mu$ being two energy scales). For $p \geq 4$, the hilltop inflation can still be compatible with current cosmological observations.

Within the framework of supergravity, the modulus field $\tau$ may be the most natural candidate for inflaton~\cite{Cicoli:2023opf}. Such models could be ``no-scale'', as the breaking scale of supersymmetry is undetermined in a first approximation, and the energy scale of the effective potential can thus be much smaller than the Planck scale~\cite{Cremmer:1983bf,Ellis:1984bm,Lahanas:1986uc,Ellis:2013xoa,Ellis:2013nxa,Romao:2017uwa,CrispimRomao:2023fij}. Recently, it has been suggested that the modulus field could also address the flavour puzzle in conjunction with modular flavour symmetries~\cite{Feruglio:2017spp,Kobayashi:2023zzc,Ding:2023htn}. The vacuum expectation value of the modulus field determining the flavour structure may be dynamically fixed by various modulus stabilisation approaches~\cite{Cvetic:1991qm, Font:1990nt, Gonzalo:2018guu,Novichkov:2022wvg, Leedom:2022zdm, King:2023snq,Ishiguro:2020tmo,Kobayashi:2019xvz, Kobayashi:2019uyt, Ishiguro:2022pde,Knapp-Perez:2023nty,Kobayashi:2023spx, Higaki:2024jdk}. Modular invariance, if applied to inflation models, can impose constraints on the corresponding scalar potential and observables~\cite{Schimmrigk:2014ica,Schimmrigk:2016bde}. It has also been noticed that modulus stabilisation and inflation can be compatible in a modular invariant framework including one additional stabiliser field and a non-minimal K\"ahler potential~\cite{Kobayashi:2016mzg,Abe:2023ylh,Gunji:2022xig}.

In this paper we show that it is possible to achieve successful hilltop inflation in which the inflaton is identified as the modulus field in a modular invariant theory. We work explicitly within the framework of modular invariant gaugino condensation~\cite{Dine:1985rz, Nilles:1982ik, Ferrara:1982qs}, which can induce the potential for modulus stabilisation~\cite{Cvetic:1991qm, Font:1990nt, Gonzalo:2018guu,Novichkov:2022wvg,Leedom:2022zdm,King:2023snq}. It was proposed in Ref.~\cite{Leedom:2022zdm} that the inclusion of non-trivial dilaton effects in this approach could uplift the potential, and thus lead to de Sitter vacua. In our work, we illustrate that the dilaton field may also play a crucial role in shaping the scalar potential and uplifting the point $\tau = \I$ to a saddle point, suitable for hilltop inflation. This is different from the other approaches to modular invariant inflation in Refs.~\cite{Kobayashi:2016mzg,Abe:2023ylh,Gunji:2022xig}. The inflationary trajectory is along the lower boundary of the fundamental domain of $\tau$, connecting two fixed points $\tau = \I$ and $\tau = \omega \equiv e^{2\I\pi/3}$, both of which are phenomenologically interesting in generating the flavour mixing structure and fermion mass hierarchy (see, e.g.,~\cite{Penedo:2018nmg, Novichkov:2018yse, Ding:2019gof,deMedeirosVarzielas:2020kji,Okada:2020ukr,Novichkov:2021evw,Feruglio:2021dte,Wang:2021mkw,Kikuchi:2022svo,Feruglio:2022koo}).
Inflation starts at a point close to $\tau = \I$, and ends before it reaches $\tau = \omega$, which is the global de Sitter vacuum. We study the slow-roll behaviour of the potential near $\tau = \I$, and investigate the allowed parameter space for successful inflationary scenario using both analytical and numerical methods.

The structure of this paper is as follows. In Sec.~\ref{sec:potential}, we briefly introduce the basic formalism of modular symmetries and construct the modular invariant scalar potential. We analyse the slow-roll behaviour of the potential near $\tau = \I$ in Sec.~\ref{sec:slowroll}. In Sec.~\ref{sec:nonminimal} we discuss an extension of the minimal framework to include a non-minimal superpotential. We summarise and conclude in Sec.~\ref{sec:sum}.

\section{Modular invariant scalar potential} \label{sec:potential}
\begin{figure}[t!]
	\centering
    \includegraphics[width=0.9\linewidth]{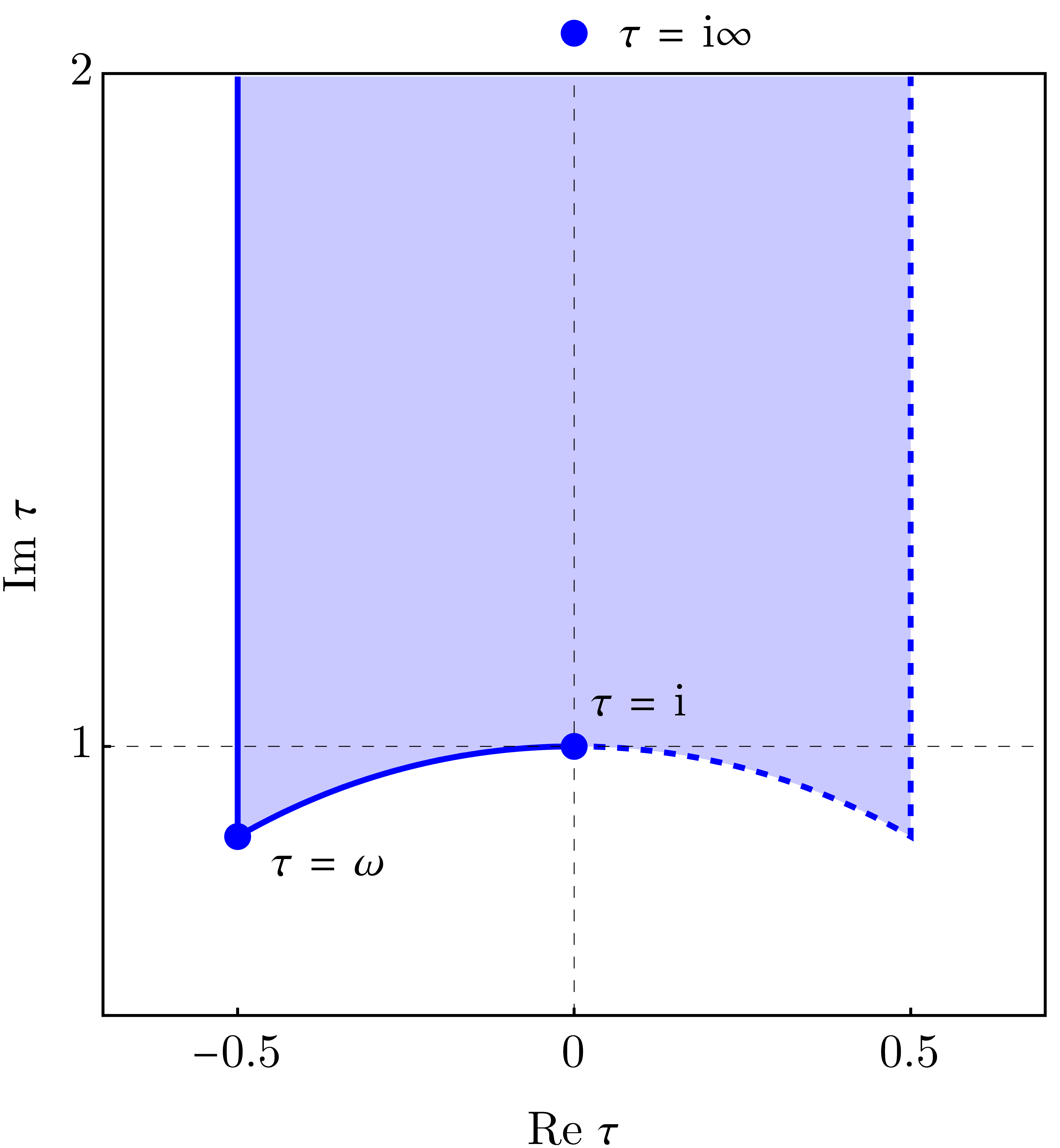}
	\caption{Fundamental domain of the modular group, where blue dots label the fixed points. The inflationary trajectory lies on the lower boundary of the fundamental domain from  $\tau = \I$  to $\tau = \omega$. }
	\label{fig:FDspec}
\end{figure}

The modulus $\tau$ can be treated as a complex field, whose value is within the upper-half complex plane $\mathbb{C}_+$, transforming under the element $\gamma$ of the modular group as
\begin{eqnarray}
\gamma: \tau \rightarrow \dfrac{a \tau + b}{c \tau + d} \; ,
\label{eq:lintran}
\end{eqnarray}
where $a, b, c, d \in \mathbb{Z}$ and $a d - b c = 1$ is satisfied. There are two generators of the modular group, namely, the duality transformation $S$: $\tau \to -1/\tau$, and the shift transformation $T$: $\tau \to \tau + 1$. The action of all elements $\gamma$ on a given point in $\mathbb{C}_+$ generates an orbit of $\tau$. Then one can find a region called fundamental domain
\begin{equation}
\begin{split}
{\cal G} = & \left\{ \tau \in \mathbb{C}^{}_{+}:  -\frac{1}{2} \leq {\rm Re}\,\tau < \frac{1}{2}, \; |\tau| > 1 \right\} \; \\
& \cup \left\{ \tau \in \mathbb{C}^{}_{+}:  -\frac{1}{2} \leq {\rm Re}\,\tau \leq 0, \; |\tau| = 1 \right\} ,
\end{split}
\label{eq:fundo}
\end{equation}
which intersects with each of these orbits in one and only one point. It provides us with the minimal space to investigate modular transformations. In~\figref{fig:FDspec} we show the fundamental domain of the modular group and the inflationary trajectory. 

We work on the ${\cal N} = 1$ supergravity model including one K\"ahler modulus $\tau$, together with one dilaton $S$. Modular invariance requires the K\"ahler function\footnote{In this paper we adopt the Planck units, i.e., the reduced Planck mass $M_{\rm Pl} = 1$.} $G(\tau, \overline{\tau}, S, \overline{S}) =  {\cal K}(\tau, \overline{\tau},S, \overline{S}) + \log \left|{\cal W}(\tau, S)\right|^2_{}$ (with ${\cal K}(\tau, \overline{\tau},S, \overline{S})$ and ${\cal W}(\tau, S)$ being respectively the K\"ahler potential and superpotential) remains unchanged under the modular transformation. Then the modular invariant scalar potential turns out to be~\cite{Cremmer:1982en}
\begin{equation}
    V=e^{\cal K}_{}\left({\cal K}^{i \bar{j}}_{} D^{}_i {\cal W} D^{}_{\bar{j}} {\cal \overline{W}}-3 |{\cal W}|^2_{}\right) \; ,
    \label{eq:scalar-potential}
\end{equation}
where the covariant derivatives $D_i \equiv \partial^{}_i + (\partial_i {\cal K})$ are defined, and ${\cal K}^{i \bar{j}}$ is the inverse of the K\"ahler metric ${\cal K}_{i\overline{j}} \equiv \partial_i \partial_{\overline{j}} {\cal K}$. Note that the subscripts $i$ and $j$ should refer to $\tau$ and $S$. 

Assuming the K\"ahler potential of $\tau$ to be minimal, the entire K\"ahler potential takes the form
\begin{equation}
{\cal K}(\tau, \overline{\tau},S, \overline{S}) =  K( S, \overline{S})-3 \log(2 \,{\rm Im}\,\tau)  \; ,
\label{eq:kahler-potential}
\end{equation}
where the K\"ahler potential for the dilaton $K( S, \overline{S}) \equiv -\ln(S+\overline{S}) + \delta K(S, \overline{S})$ with $\delta K(S, \overline{S})$ being additional corrections from some stringy effects, e.g., Shenker-like effects~\cite{Shenker:1990}, which would be necessary for the dilaton stabilisation~\cite{Leedom:2022zdm}.

In general, moduli originate from orbifold compactifications. In the simplest supersymmetric compactification, the potential of moduli remains flat. The inclusion of non-perturbative effects may uplift the potential and thus stabilise the moduli fields. Here we especially focus on the gaugino condensation mechanism in the heterotic string~\cite{Dine:1985rz, Nilles:1982ik, Ferrara:1982qs}. The superpotential can in general be parametrised as~\cite{Cvetic:1991qm}
\begin{equation}
    {\cal W}(\tau,S) = {\Lambda^3_W}\frac{\Omega(S)H(\tau)}{\eta^6(\tau)} \; ,
    \label{eq:superp-para}
\end{equation}
where $\Lambda_W$ labels the energy scale (in units of $M_{\rm Pl}$) of gaugino condensation, $\Omega(S)$ denotes the contribution from dilaton sector, and $\eta(\tau)$ is the Dedekind $\eta$-function. The dependence of ${\cal W}(\tau,S)$ on $\eta^{-6}(\tau)$ is induced by the one-loop threshold corrections~\cite{Giddings:2001yu, Gukov:1999ya, Curio:2000sc, Ashok:2003gk, Denef:2004cf}. As a result, ${\cal W} \to (c\tau + d)^{-3} {\cal W}$ under the modular transformation $\tau \to \gamma \tau$. Then one can identify that the modular transformations of ${\cal K}$ and ${\cal W}$ compensate with each other, leading to a modular-invariant $G$. In addition, $H(\tau)$ is a dimensionless function that parametrises higher-order corrections to the superpotential. To maintain the modular invariance and avoid any singularity of $H(\tau)$ inside the fundamental domain,  $H(\tau)$ should be a rational function in terms of the modular invariant Klein $j$-function $j(\tau)$. Then we have~\cite{Cvetic:1991qm}
\begin{equation}
    H(\tau) = (j(\tau)-1728)^{m/2}j(\tau)^{n/3}{\cal P}(j(\tau)) \; ,
    \label{eq:H-def}
\end{equation}
with $m$ and $n$ being non-negative integers, and ${\cal P}(j(\tau))$ representing a polynomial of $j(\tau)$. The definitions of $\eta(\tau)$ and $j(\tau)$ are given in Appendix~\ref{app:modular-forms}. 

\begin{figure}[t!]
	\centering
    \includegraphics[width=0.9\linewidth]{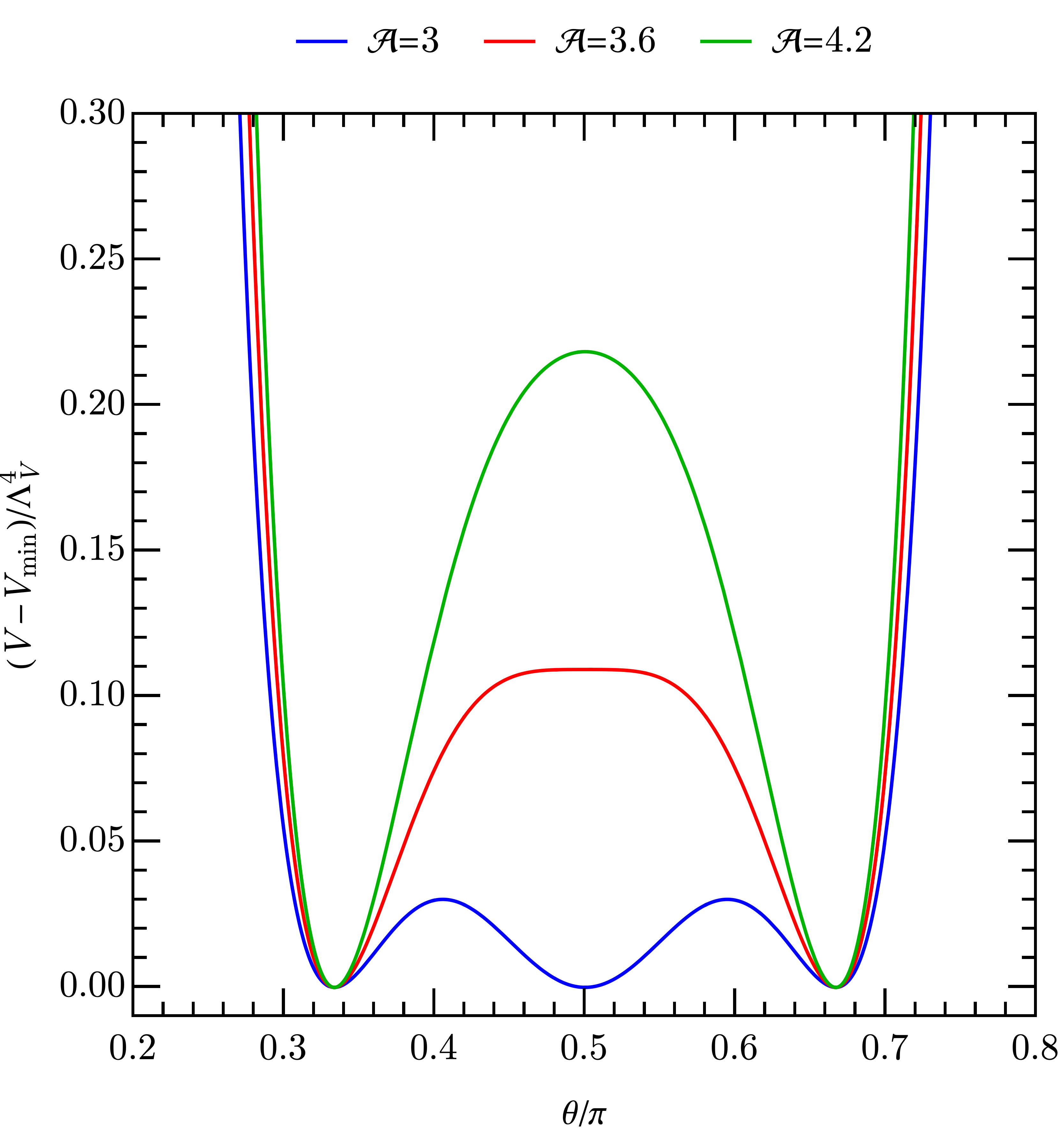}
	\caption{The evolution of $(V - V_{\rm min})$ with $V_{\rm min}$ being the minimum of the scalar potential along the lower boundary of the fundamental domain parametrised by $\tau = e^{\I \theta}$. Blue, red and green curves correspond to ${\cal A} = 3$, 3.6 and 4.2, respectively. With the increase of ${\cal A}$, the fixed point $\tau = {\rm i}$ transitions from a local minimum to a local maximum in the $\theta$-direction.}
	\label{fig:potvar}
\end{figure}

As a very simple scenario, we assume $H(\tau) = 1$, indicating only the one-loop threshold corrections are taken into consideration. Then the scalar potential is explicitly expressed as~\cite{Leedom:2022zdm}
\begin{equation}
    \begin{split}
    V=& \dfrac{\Lambda^4_V e^{K(S,\overline{S})}_{}|\Omega(S)|^2_{}}{(2\,{\rm Im}\,\tau)^3_{}|\eta(\tau)|^{12}_{}}\bigg[\frac{(2\,{\rm Im}\,\tau)^2_{}}{3}\left|\frac{3}{2\pi}\widehat{G}^{}_2(\tau,\overline{\tau})\right|^2_{}  \\
    &+\left({\cal A}(S, \overline{S}) - 3 \right) \bigg] \; ,
    \end{split}
    \label{eq:single-ponten}
\end{equation}
where the energy scale $\Lambda^{}_V = \Lambda^{3/2}_W$ and $\widehat{G}_2$ is the non-holomorphic Eisenstein series of weight 2 (see appendix~\ref{app:modular-forms} for definition). $\calAS$ depends on $S$ and $\overline{S}$ in the form
\begin{equation}
    \mathcal{A}(S,\overline{S}) = \frac{|\Omega^{}_S+K^{}_S\Omega|^2_{}}{K^{}_{S\overline{S}}|\Omega|^2_{}} \; ,
    \label{eq:AS}
\end{equation}
where the subscripts represent the derivatives with respect to $S$ or $\overline{S}$. If $K(S,\overline{S})$ takes the minimal form, i.e, $ K(S,\overline{S}) = \ln(S+\overline{S})$, $\Omega^{}_S+K^{}_S\Omega$ must be vanishing to guarantee the scalar potential is also stable in the dilaton sector. Nevertheless, it was proposed in 
Ref.~\cite{Leedom:2022zdm} that if additional terms $\delta K(S, \overline{S})$ from Shenker-like effects are introduced into $K(S,\overline{S})$, it is possible to stabilise the potential in the dilaton sector even if $\mathcal{A}(S,\overline{S}) \propto |\Omega^{}_S+K^{}_S\Omega|^2 \neq 0$. 

For the scalar potential in Eq.~(\ref{eq:single-ponten}), it is known $\tau = \omega$ always corresponds to a de Sitter minimum as long as ${\cal A}(S,\overline{S}) > 3$~\cite{Leedom:2022zdm,King:2023snq}, while $\tau = {\rm i}$ can only be a minimum when $3 \leq {\cal A}(S,\overline{S}) \leq 3.596$. When ${\cal A}(S,\overline{S}) > 3.596$, $\tau = {\rm i}$ becomes a saddle point. In~\figref{fig:potvar}, we plot the evolution of $(V - V_{\rm min})$ along the lower boundary (parametrised by $\theta$) with different choices of $\calA(S,\overline{S})$. The value of $\calA(S,\overline{S})$ dramatically adjusts the shape of $V$. In particular, the potential becomes fairly flat near $\tau = {\rm i}$ (or equivalently $\theta = \pi/2$) when $\calA(S,\overline{S}) \simeq 3.596$. Hence we expect the slow-roll inflation might occur, with the modulus $\tau$ initialised at a point close to $\tau = {\rm i}$, and rolling down to another fixed point $\tau = \omega$, which is the true vacuum of the potential. It is interesting to ask why the initial value of $\tau$ can be close to $\tau = \I$. One possible way is to consider the dynamical evolution of $\calA(S,\overline{S})$. Then the modulus may be at the vacuum $\tau = \I$ at an earlier time, which is uplifted later as the potential in the vicinity of $\tau =\I$ becomes flat. This process and the dilaton stabilisation may be carried out simultaneously, the details of which are left for future work. Here we just assume the scalar potential has already been stabilised in the dilaton sector due to the existence of $\delta K(S,\overline{S})$, but do not focus on its explicit form. Then ${\cal A}$ is treated as a free parameter, and $e^{K(S,\overline{S})}|\Omega(S)|^2$ in Eq.~(\ref{eq:single-ponten}) is absorbed into the overall energy scale $\Lambda_V$. 

\begin{figure*}[t!]
	\centering
\includegraphics[width=0.96\linewidth]{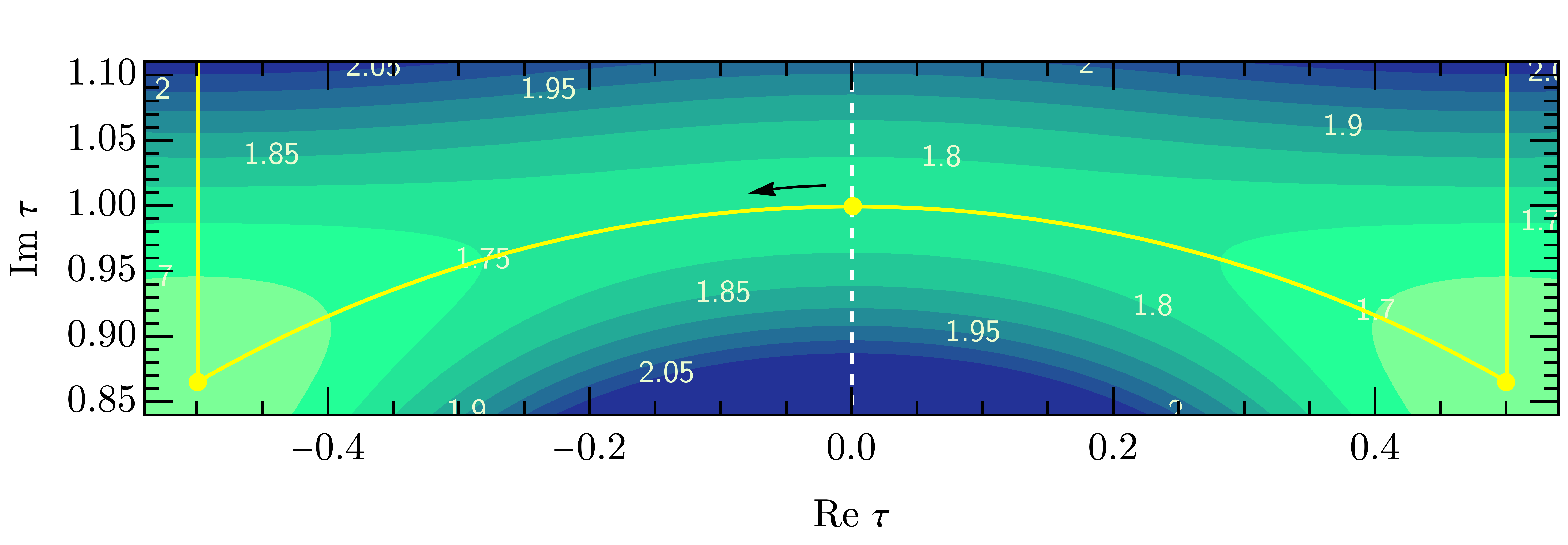}
	\caption{Contour plot of the potential $V$ in the ${\rm Re}\,\tau - {\rm Im}\,\tau$ plane in the case where $\calA = 3.6$. Numbers on the contours correspond to the values of $V$ in units of $\Lambda_V^4$. The yellow lines outline the boundaries of fundamental domain. The arrow labels the direction of the slowly-rolling inflaton trajectory.}
	\label{fig:traject}
\end{figure*}

On the other hand, the modulus field is a complex scalar with two degrees of freedom. In \figref{fig:traject}, we show the contour plot of the potential $V$ in the ${\rm Re}\,\tau - {\rm Im}\,\tau$ plane by choosing $\calA = 3.6$, where we can observe that each point of the lower boundary is the local minimum in the direction perpendicular to the $\theta$-direction. Then it is reasonable to consider only single-field inflation if the modulus initially lies on the lower boundary of the fundamental domain.

\section{Slow-roll behaviour near $\tau = {\rm i}$} \label{sec:slowroll}
Now we start to investigate the slow-roll behaviour of the modulus field near $\tau = {\rm i}$. It is more convenient to divide $\tau$ into the radial component $t$ and the angular component $\theta$, i.e., $\tau = t e^{{\rm i}\theta}$. Then the kinetic Lagrangian of $\tau$ can be expressed as
\begin{equation}
    {\cal L}_{\rm kin} =3\left(\partial_\mu t ~ \partial_\mu \theta \right) 
    \left(
    \begin{matrix}
        t^{-2} & \\
        & \csc^2\theta 
    \end{matrix}
    \right)
    \left( 
    \begin{matrix}
        \partial^\mu t \\
        \partial^\mu \theta
    \end{matrix}\right) \; .
\end{equation}
In order to make $ {\cal L}_{\rm kin} $ canonical, we should convert $t$ and $\theta$ to two new real fields $\rho$ and $x$ by $t \equiv e^{-\rho/\sqrt{3}}$ and $\theta \equiv 2\,{\rm arctan} (e^{x/\sqrt{3}})$. 
As a result, the Lagrangian becomes ${\cal L} = (\partial_\mu \rho \partial^\mu \rho + \partial_\mu x \partial^\mu x)/2 -V(\rho, x)$. Since we mainly care about the inflationary trajectory along the lower boundary, $\rho = 0$ is fixed. We can implement the series expansion of the scalar potential $V(x)$ near $x=0$ (which corresponds to $\tau = \I$). Up to the fourth order of $x$, we have the following form (see Appendix~\ref{app:derivation} for the derivation)
\begin{widetext}
\begin{eqnarray}
    \dfrac{V(x)}{\Lambda^4_V} &= &  \dfrac{512\pi^9({\cal A}-3)}{\Gamma^{12}(1/4)} + \dfrac{\pi(192\pi^4-\Gamma^8(1/4)) ( 192({\cal A}-2)\pi^4 - \Gamma^8(1/4) )}{ {288\Gamma^{12}(1/4)}}x^2   + \left[ \left(\dfrac{128\pi^9}{9\Gamma^{12}(1/4)} -\dfrac{7\pi^5}{54\Gamma^4(1/4)}\right.\right. \nonumber \\
    &&  \left.\left. +\dfrac{\pi\Gamma^{4}(1/4)}{1152} \right){\cal A}\right.- \left. \left( \dfrac{160\pi^{9}}{9\Gamma^{12}(1/4) } - \dfrac{2\pi^5}{27\Gamma^4(1/4)} - \dfrac{5\pi\Gamma^4(1/4)}{5184} + \dfrac{\Gamma^{12}(1/4)}{73728\pi^3} \right) \right] x^4 + {\cal O}(x^6) \; ,
    \label{eq:scalar-expand}
\end{eqnarray}
\end{widetext}
where one can recognise the terms with odd powers of $x$ automatically vanish, which can be understood since the $V(x)$ is a symmetric function regarding the axis $x = 0$. One can further allow the quadratic term of $V(x)$ to be zero by requiring
\begin{equation}
    {\cal A} = 2 + \frac{\Gamma^8(1/4)}{192\pi^4} \approx 3.596 \; .
    \label{eq:2cancelcond}
\end{equation}
It is interesting to mention that the elimination of the quadratic coupling also means the vanishing of the Hessian. Therefore, Eq.~(\ref{eq:2cancelcond}) indeed indicates where the fixed point $\tau = \I$ transitions from a local minimum to a local maximum in the $x$ direction, in accordance with the results in Refs.~\cite{Leedom:2022zdm, King:2023snq}. 
Then the scalar potential approximates to
\begin{equation}
    V(x) \approx \widetilde{\Lambda}^4_V [1 - 0.306 x^4 + {\cal O}(x^6) ] \; ,
    \label{eq:potenapprox}
\end{equation}
where $\widetilde{\Lambda}^4_V \equiv 1.764 \Lambda^4_V$. The above equation precisely indicates a $p = 4$ hilltop-like scalar potential. However, with a large quartic coefficients, this simple potential does not agree with the latest experimental observations~\cite{Planck:2018jri}.

There are two important dimensionless parameters that can describe the slow-roll inflation, namely,\footnote{Following the commonly-used convention, we use $\eta$ to denote the ratio of $V^{\prime\prime}$ and $V$, which should be distinguished from the Dedekind $\eta$-function $\eta(x)$ in the present paper.}
\begin{equation}
    \epsilon \equiv \frac{1}{2}\left(\frac{V^\prime}{V}\right)^2 \; , \quad \eta \equiv  \frac{V^{\prime\prime}}{V} \; .
    \label{eq:deriva}
\end{equation}
With the help of $\epsilon$ and $\eta$, the tensor-to-scalar ratio $r$ and the spectral index $n_s$ can be respectively estimated as
\begin{equation}
    r \simeq 16\epsilon \; , 
    \quad
    n_s \simeq 1 - 6 \epsilon + 2 \eta \; .
    \label{eq:para-def}
\end{equation}
        
Apart from $n_s$ and $r$, there is another observable, which is the scalar amplitude $A_s$. For a certain scalar potential, $A_s$ can be calculated by
\begin{equation}
    A_s = \frac{1}{24\pi}\frac{V}{\epsilon} \; .
    \label{eq:scalar-amplitude}
\end{equation}
In fact,  $A_s$ is the only observable that can determine the overall scale of the scalar potential.

The scale factor should be dramatically increasing during the inflation epoch, which can be evaluated by the e-folds number
\begin{equation}
    N_e(x) = \int_{x}^{x_e} \frac{1}{\sqrt{2\epsilon}} {\rm d}x \; ,
    \label{eq:efold-exp}
\end{equation}
where $x_e$ denotes the field value at the end of inflation. Successful inflationary scenario prefers $50 \lesssim N_e \lesssim 60$, with which we can determine the field value $x_i$ at the beginning of inflation. Then we can calculate $n_s$ and $r$ correspondingly.

Let us first make some analytical predictions using the approximate scalar potential. If Eq.~(\ref{eq:2cancelcond}) is not strictly satisfied, the scalar potential can be recast into
\begin{equation}
    V(x) \approx \widetilde{\Lambda}^4_V [1 -0.25 \mathfrak{a} x^2 - (0.306 - 0.697\mathfrak{a} ) x^4 + {\cal O}(x^6)] \; ,
    \label{eq:poten-a}
\end{equation}
where $\mathfrak{a} \equiv \calA - 3.596$ is a small parameter. Using Eq.~(\ref{eq:deriva}), we arrive at
\begin{eqnarray}
   \epsilon & \approx & 0.750 x^6 + (0.612 x^4 -3.413 x^6)\mathfrak{a}   \; , \label{eq:eps-appro} \\
   \eta & \approx & -(3.674x^2-11.685x^4) - (0.5 - 8.362x^2)\mathfrak{a} \; ,
\label{eq:xi-appro}
\end{eqnarray}
Substituting Eq.~(\ref{eq:eps-appro}) into Eq.~(\ref{eq:efold-exp}), the e-folds number $N_e(x)$ can be evaluated as
\begin{equation}
    N_e(x) \approx \left. -\frac{1}{\mathfrak a}\log\left( 1 + \frac{0.408 - 2.276 x^2}{x^2}\mathfrak{a}\right) \right|^{x_e}_{x}\; .
    \label{eq:Ne-appro}
\end{equation}
For the scalar potential of our interest, $\epsilon \ll |\eta|$ is always satisfied. Therefore the slow-roll inflation should end when $|\eta| \approx 1$. On the other hand, the function in the right-hand side of Eq.~(\ref{eq:Ne-appro}) takes the limit $2.276 + 2.590\mathfrak{a}$ as $x$ tends to $+\infty$, which can be regarded as the e-folds number at the end of inflation. Therefore, from Eq.~(\ref{eq:Ne-appro}) we can solve out $x_i$ with a given e-folds number $N_e$ during inflation, namely,
\begin{equation}
    x^{}_i \approx \frac{0.082}{\widehat{N}_e}+\left(\frac{0.093}{\widehat{N}_e}-1.237\widehat{N}_e\right)\mathfrak{a} + 3.093n^3_e \mathfrak{a}^2_{}\; ,
\label{eq:xini}
\end{equation}
where $\widehat{N}_e \equiv \sqrt{N_e/60}$ is of ${\cal O}(1)$ when $50 \lesssim N_e \lesssim 60$. Combining the above equation with Eqs.~(\ref{eq:para-def}), (\ref{eq:eps-appro}) and (\ref{eq:xi-appro}), we can establish the following relations among $\widehat{N}_e$, $r$ and $n_s$
\begin{eqnarray}
    r &\approx& \left(\frac{3.78 }{\widehat{N}_e^6}  + \frac{113}{\widehat{N}^4_e}\mathfrak{a} - \frac{851}{\widehat{N}_e^2}\mathfrak{a}^2_{} \right) \times 10^{-6}_{} \; , \label{eq:r-N}\\
    n_s^{} & \approx &  1 - \frac{0.05}{\widehat{N}^2_e} + 0.5\mathfrak{a} - 11.25\widehat{N}^2_e\mathfrak{a}^2_{} \; .
    \label{eq:ns-Ne}
\end{eqnarray}
From Eq.~(\ref{eq:r-N}), one could recognise $r \sim 10^{-6}$ is negligibly small. On the other hand, if we require $\widehat{N}_e \lesssim 1$ (or equivalently $N_e \lesssim 60$), $n_s$ should be smaller than $0.950 + 0.5\mathfrak{a} -11.25\mathfrak{a}^2$, the maximal value of which turns out to be 0.956 at $\mathfrak{a} = 0.022$, i.e., $\calA = 3.618$. 

\begin{figure}[t!]
	\centering
\includegraphics[width=1\linewidth]{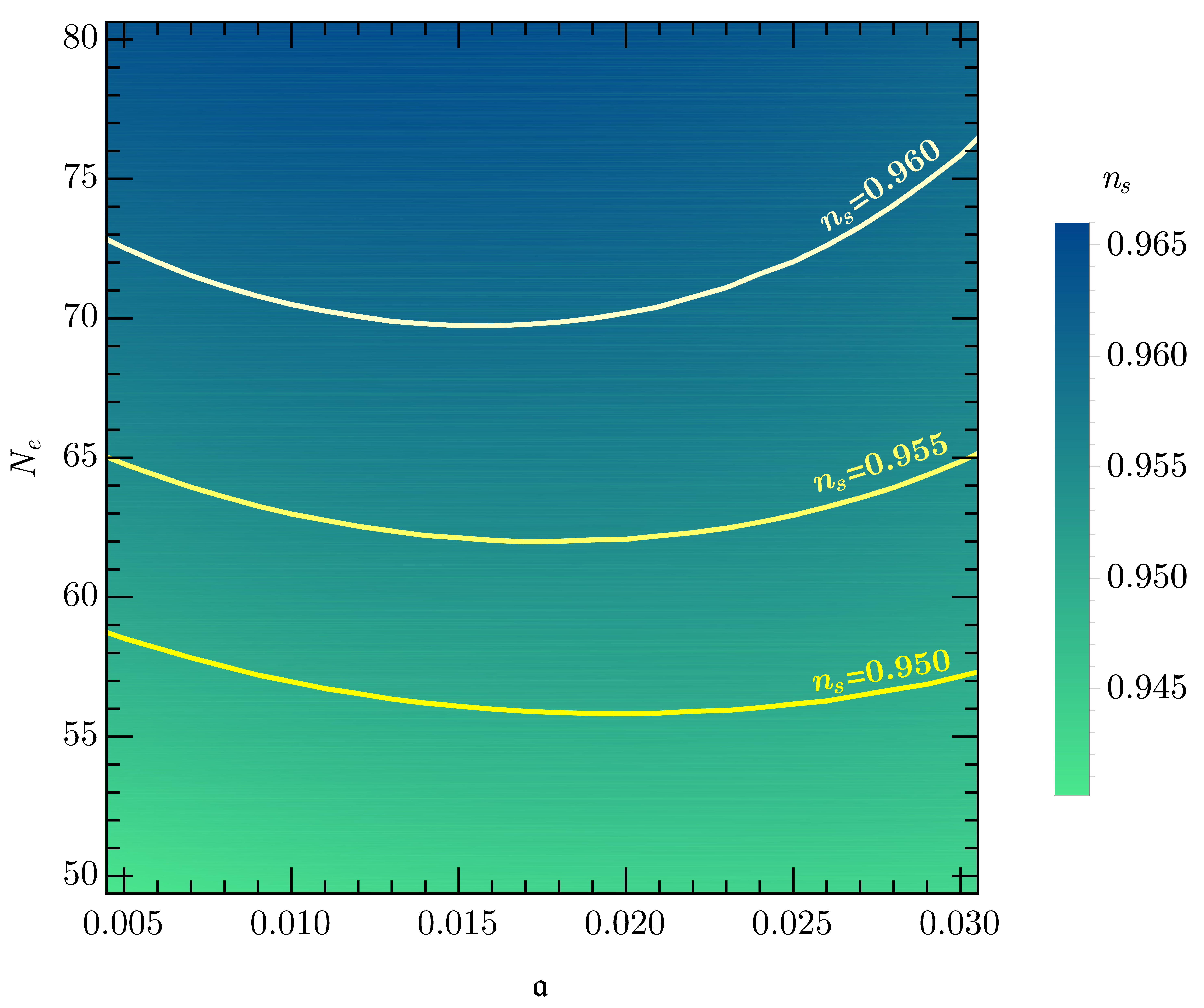}
	\caption{Distribution of the spectral index $n_s$ with varying $\mathfa$ and $N_e$. The curves correspond to $n_s = 0.960$, 0.955 and 0.950, respectively. The upper curve is slightly below the experimental $1\sigma$ range $0.9621 < n_s < 0.9701$.}
	\label{fig:nsvalue}
\end{figure}

In~\figref{fig:nsvalue}, we exhibit the distribution of $n_s$ with varying $\mathfa$ and $N_e$, where we see that $n_s$ can not be greater than 0.955 if $50 < N_e < 60$ is required. This is in agreement with the analytical calculation. As a result, the model with $H(\tau)=1$ in the first approximation is slightly disfavoured by the experimental observations.

\section{Modifications from non-minimal superpotential} \label{sec:nonminimal}

The mild tension between our predictions and experimental measurements on inflationary observables arises from the fact that changing the value of $\mathfa$ does not make the quadratic and quartic coefficients in Eq.~\eqref{eq:poten-a} sufficiently small. In order to provide a better fit to the experimental observations, one may introduce modifications to $H(\tau)$ without violating its modular invariance. It is found that the appropriate form of $H(\tau)$ that can realise successful inflation should be $H(\tau) = 1 + \alpha \delta(\tau)$, where $\delta(\tau) \equiv [j(\tau)/1728 - 1]^2$, and $\alpha$ is a small parameter.\footnote{We can also find allowed parameter space for successful inflation with a linear modification $\delta(\tau) = [j(\tau)/1728-1]$. However in such a case, we would need $\mathfa < 0$, indicating $\partial^2 V/\partial x^2 > 0$ at $x=0$ without modification, i.e., the leading order potential with $H(\tau)=1$ would correspond to a local minimum at $\tau = \I$. Moreover in such a case the magnitude of $\mathfa$ would be larger. Instead, we prefer to consider the leading order case with $H(\tau)=1$ in the first approximation, which can already give qualitatively correct slow-roll behaviour. 
Then only a small quadratic correction $\delta(\tau) = [j(\tau)/1728 - 1]^2$ is required.}
Then the scalar potential turns out to be 
\begin{equation}
\begin{split}
    V(\tau, \overline{\tau}) = & \frac{{\Lambda_V^4}}{(2\,{\rm Im}\,\tau)^3_{}|\eta(\tau)|^{12}_{}}\left[\frac{(2\,{\rm Im}\,\tau)^2_{}}{3}\bigg|{\rm i}\alpha \delta^\prime \right.\\
    & +\left. \frac{3(1 + \alpha \delta)}{2\pi}\widehat{G}^{}_2(\tau,\overline{\tau})\bigg|^2_{} + \left({\cal A} - 3 \right)|1 + \alpha \delta|^2 \right] \; .
    \end{split}
    \label{eq:single-ponten-2}
\end{equation}
\begin{figure}[t!]
	\centering
	\includegraphics[width=0.9\linewidth]{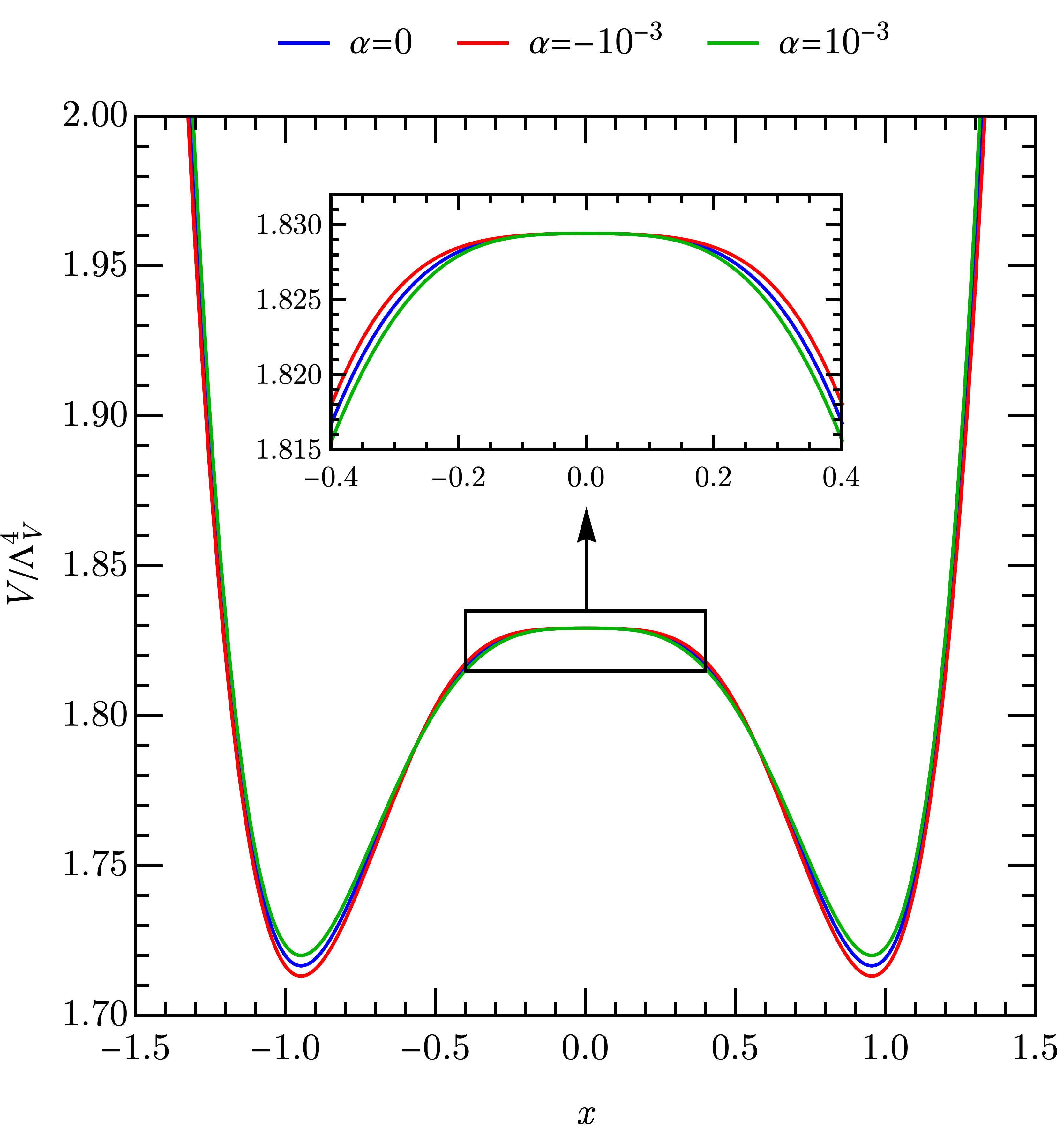}
	\caption{The shape of scalar potential with different values of $\alpha$. A negative $\alpha$ makes the potential  near $\tau = \I$ flatter. }
	\label{fig:potennm}
\end{figure}

After the field redefinition, in the vicinity of $x=0$ the potential approximates to
\begin{equation}
    V(x) \approx \widetilde{\Lambda}^4_V [ 1- 0.25\mathfrak{a} x^2_{}  - f^{}_4(\mathfa,\alpha)x^4_{} +f^{}_6(\mathfa, \alpha)x^{6}_{} + {\cal O}(x^{8}_{})] \; ,
\end{equation}
where 
\begin{equation}
\begin{split}
    f^{}_4(\mathfa,\alpha) \equiv & (0.306 - 0.697\mathfa) + (138 - 308\mathfa)\alpha \; , \\
    f^{}_6(\mathfa,\alpha)  \equiv & (0.389 - 0.702\mathfa) + (1231 - 2397\mathfa)\alpha \\
    & + (56455 -94656\mathfa)\alpha^2_{} \; .
\end{split}
\end{equation}
The quadratic coefficient $-0.25\mathfa$ is always negative as $\mathfa>0$, while the coefficients $-f^{}_4(\mathfa,\alpha)$ and $f^{}_6(\mathfa,\alpha)$ could change their signs in the presence of $\alpha$. Then it is possible to realise successful inflation by making $f^{}_6(\mathfa,\alpha)$ negative, which leads us to the $p = 6$ hilltop inflation. This requires $\alpha$ to be within the range $-(21.5 - 0.60\mathfa) < \alpha/10^{-3} < -(3.21 + 0.49\mathfa)$, which is around ${\cal O}(10^{-2} \cdots 10^{-3})$ for small $\mathfa$. In \figref{fig:potennm}, we illustrate the influence of different values of $\alpha$ on the scalar potential. Negative $\alpha$ indeed makes the potential even flatter around $x=0$. We can also check the validity of this scenario in a more precise way with the help of analytical approximations. Here we choose $\mathfa = 0.022$ for illustration. Neglecting higher-order terms of $x$, the expression for $N_e$ up to ${\cal O}(\alpha)$ reads
\begin{equation}
\begin{split}
    N_e(x)  = & \int_{x}^{x_e} \frac{1}{\sqrt{2\epsilon}} {\rm d}x  \approx \int_{x}^{x_e} \left[\frac{0.860}{xg(x)} + \frac{387x}{g(x)^2}\alpha \right] {\rm d}x \\
    = & \left.\left[ 45.4\log\left(\frac{x^2}{g(x)}\right) + \frac{193}{g(x)}\alpha\right]\right|^{x_e}_x \; ,
\end{split}
\label{eq:efold-exp-nm}
\end{equation}
where $g(x) \equiv x^2 + 0.00946$. Meanwhile, the approximated expression for $\eta$ becomes
\begin{eqnarray}
    \eta &\approx& -(0.011+3.49x^2-11.7x^4)-(1570x^2-36920x^4)\alpha \nonumber \\
    && + 169365 x^4 \alpha^2 \; ,
    \label{eq:eta2}
\end{eqnarray}
which reduces to Eq.~(\ref{eq:xi-appro}) with $\mathfa = 0.022$ as $\alpha$ tends to $0$. Since the field value $x^{}_e$ at the end of inflation should be large, the $x^4_{}$-terms in Eq.~(\ref{eq:eta2}) would dominate the value of $\eta$. If we further require $\alpha \simeq {\cal O}(10^{-3}_{})$, $x^{}_e$ can be roughly evaluated via $|\eta| \simeq 36920x^4_e|\alpha| \simeq 1$, leading to $x^{}_e \simeq 0.073|\alpha|^{-1/4}_{} \simeq 0.4$. Assuming the inflation starts at $x_i = 0.08$, from Eq.~(\ref{eq:efold-exp-nm}) we arrive at $\alpha \simeq -(6.82\widehat{N}_e-3.14)\times 10^{-3}$. Using Eqs.~(\ref{eq:para-def}) and (\ref{eq:eta2}), one can obtain $n_s \approx 1 + 2\eta \approx 0.876 + 0.08\widehat{N}_e +0.004\widehat{N}_e^2$, which turns out to be 0.968 when $\widehat{N}_e = 1$. Hence the inclusion of the $\alpha \delta(\tau)$ term in the superpotential indeed enhances the consistency between our predictions and the observational results.
\begin{figure}[t!]
	\centering
	\includegraphics[width=0.9\linewidth]{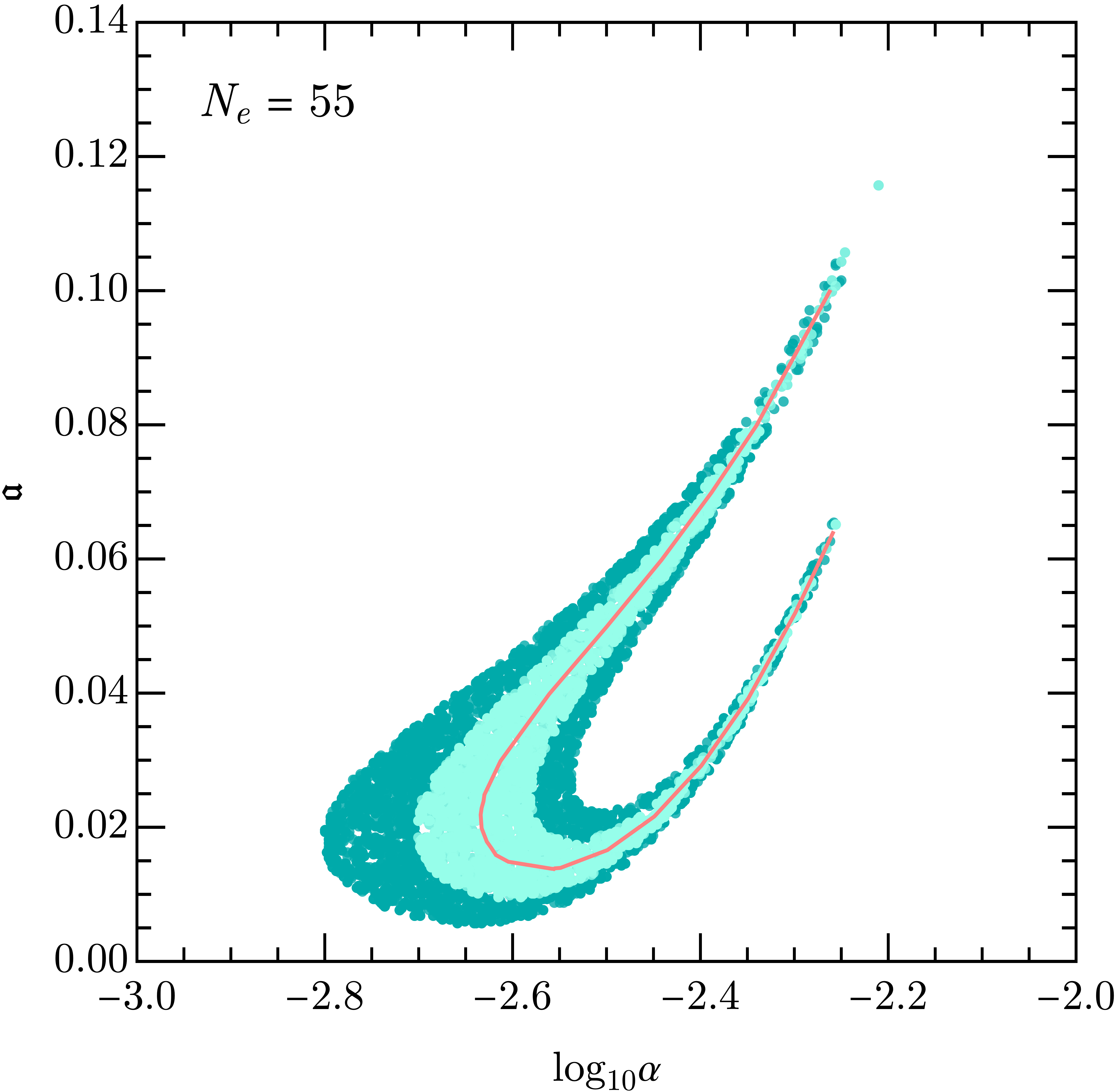}
	\caption{The allowed parameter space in the $\log_{10}\alpha - \mathfa$ plane that is consistent with the current experimental constraint on $n_s$, where $N_e = 55$ is assumed. The light and dark green shaded regions denote the $1\sigma$ (68\% CL) and $2\sigma$ (95\% CL) ranges, respectively. The pink curve corresponds to the best-fit value of $n_s$.}
	\label{fig:paraspace}
\end{figure}

We perform a $\chi^2$-analysis to obtain the feasible parameter space of $\alpha$ and $\mathfa$. In our analysis, we fix $N_e = 55$. Then there are two inflationary observables that can determine the allowed ranges of $\alpha$ and $\mathfa$, namely, $r$ and $n_s$. In our model, $r \simeq 10^{-6}$ is considerably lower that its experimental bound. Hence $n_s$ becomes the only observable relevant for the $\chi^2$-function
\begin{eqnarray}
    \chi^2_{}(\alpha,\mathfa) = \frac{(n^{}_s-n_s^{\rm bf})^2}{\sigma^2_{n_s}} \; ,
    \label{eq:chi2}
\end{eqnarray}
with $n_s^{\rm bf} = 0.9661$ and $\sigma^{}_{n_s} = 0.0040$. In Fig.~\ref{fig:paraspace}, we show the $1\sigma$ (68\% CL) and $2\sigma$ (95\% CL) ranges of $\alpha$ and $\mathfa$, with light and dark green shaded colours, respectively. In addition, the best-fit values of model parameters are  labelled by the pink curve in Fig.~\ref{fig:paraspace}. The results are in accordance with our analytical estimation.

Finally let us determine the overall energy scale of the potential. Substituting one best-fit point with $\alpha = - 2.73 \times 10^{-3}$ and $\mathfa = 0.040$ into Eq.~\eqref{eq:scalar-amplitude}, and taking the best-fit value of $A_s$ from the Planck 2018 TT+lowE result, namely, ${\rm ln}(10^{10}_{}A^{}_s) = 3.040$, we can get $\Lambda_V \simeq 5.7\times 10^{-4}_{}M^{}_{\rm Pl}$, which means the gaugino condensation scale $\Lambda^{}_W = \Lambda_V^{2/3} \simeq 6.9 \times 10^{-3}_{}M^{}_{\rm Pl}$. This is a very interesting scale for gaugino condensation, as it may naturally solve the puzzles of mass scales and hierarchies in physics~\cite{Nilles:1982ik}. On the other hand, it will give very large contribution to the cosmological constant. which must be cancelled somehow after inflation. This is the usual cosmological constant problem. Here we provide a possible way to resolve this problem by introducing a ``waterfall field''. Such a field adds 
another direction to the scalar potential. If falling into this direction becomes preferred at a certain point of the inflationary trajectory near $\tau = \omega$, the inflationary era will end immediately and the theory rapidly reaches another minimum with a lower energy scale~\cite{Antoniadis:2020stf}. In this case, the fixed point $\tau = \omega$ is actually an unstable de Sitter vacuum. We speculate that the waterfall field could be another modulus field in a multi-moduli scenario. An illustrative example is given in appendix~\ref{app:waterfall}.

\section{Summary and conclusion} \label{sec:sum}
In this paper we have shown that it is possible to achieve successful hilltop inflation where the inflaton is identified as the modulus field in a modular invariant theory. We have seen that 
the dilaton plays a crucial role in shaping the potential, and 
modular invariant gaugino condensation provides the mechanism for the modulus stabilisation after inflation.

Our approach includes modular invariant gaugino condensation in the heterotic string, associated with one K\"ahler modulus together with one dilaton. Such a framework has been extensively investigated for modulus stabilisation in previous literature. As we have briefly reviewed in Sec.~\ref{sec:potential}, imposing modular invariance on gaugino condensation, one can derive the superpotential as given in Eq.~(\ref{eq:superp-para}), which together with the certain form of K\"ahler potential, will finally lead to the non-trivial scalar potential [cf. Eq.~(\ref{eq:single-ponten})]. In this work, we point out that the above scalar potential may not only address the modulus stabilisation problem, but also lead to successful inflation. 

The key point is that the scalar potential depends non-trivially on the dilaton through the term $\calAS$. As $\calAS$ gradually increases starting from 3, the fixed point $\tau = \I$ evolves from a local minimum to a saddle point. Then the scalar potential near $\tau = \I$ becomes rather flat, allowing hilltop inflation to occur. The inflationary trajectory lies on the lower boundary of the fundamental domain of $\tau$, where the inflation starts near the fixed point $\tau =\I$, and ends at a point near $\tau = \omega$, which is the global de Sitter vacuum.
We have analysed the slow-roll behaviour of the potential near $\tau = \I$, and investigated the allowed parameter space for successful modular invariant hilltop inflation. We find that the minimal case predicts values for the spectral index just outside the $1\sigma$ range of current experiments.

We have also considered a modification of $H(\tau)$ by $\alpha[j(\tau)/1728 - 1]^2$ to provide a perfect fit to the spectral index. We have obtained approximate analytical relations among inflationary observables, while adopting numerical calculations to obtain the allowed parameter space. It is worth mentioning that we do not consider two-field inflation or the dynamical evolution of the dilaton sector, which are beyond the scope of the present paper but could form the basis of a future study.

\vspace{0.5cm}

\noindent {\bf Note added}: We note that a related work \refref{Ding:2024neh} appeared on the arXiv one day before the present paper was submitted.

\acknowledgments
We would like to thank George Leonaris for useful conversations at Mainz, and Ye-Ling Zhou for helpful discussions. We also acknowledge Wenbin Zhao for pointing out the possibility of having a linear correction to $H(\tau)$. SFK acknowledges the STFC Consolidated Grant ST/L000296/1 and the European Union's Horizon 2020 Research and Innovation programme under Marie Sklodowska-Curie grant agreement HIDDeN European ITN project (H2020-MSCA-ITN-2019//860881-HIDDeN). XW acknowledges the Royal Society as the funding source of the Newton International Fellowship. The authors would like to express special thanks to the Mainz Institute for Theoretical Physics (MITP) of the Cluster of Excellence PRISMA+(Project ID 390831469), for its hospitality and support.

\appendix
\section{Dedekind $\eta$-function, Eisenstein series and Klein $j$-function}\label{app:modular-forms}
The Dedekind $\eta$-function is a modular form with a weight of $-1/2$ defined as
\begin{equation}
 \eta(\tau)\equiv q^{1 / 24} \prod_{n=1}^{\infty}\left(1-q^n\right) \; ,
\label{eq:dedeta}
\end{equation}
where $q = e^{2{\rm i}\pi \tau}$.
The Eisenstein series $G^{}_{2k}(\tau)$ ($k>1$) are holomorphic modular forms with weights of $2k$, the definitions of which are
\begin{equation}
    G_{2 k}(\tau)=\sum_{\substack{n_1, n_2 \in \mathbb{Z} \\\left(n_1, n_2\right) \neq(0,0)}}\left(n_1+n_2 \tau\right)^{-2 k} \; .
    \label{eq:eisenstein}
\end{equation}
Notice that one can still define the Eisenstein series $G^{}_2(\tau)$ via a specific prescription on the order of the above summation, which is however not a modular form. Instead, one can define a non-holomorphic Eisenstein series of weight two as
\begin{equation}
    \widehat{G}_2(\tau) = G_2(\tau) - \frac{\pi}{{\rm Im}\,\tau} \; .
    \label{eq:nonholoG2}
\end{equation}
The Fourier expansions of $G_{2k}(\tau)$ are found to be
\begin{equation}
G_{2 k}(\tau)=2 \zeta(2 k)\left(1+c_{2k}\sum_{n=1}^{\infty} \sigma_{2 k-1}(n) q^n\right) \; ,
\label{eq:eisen-fourier}
\end{equation}
where $c_{2k} = 2/\zeta(1-2k)$ with $\zeta(z)$ being the  Riemann $\zeta$-function. $\sigma_a(n)$ is the divisor sum function, and we have the following substitution relation
\begin{equation}
\sum_{n=1}^{\infty} q^n \sigma_a(n)=\sum_{n=1}^{\infty} \frac{n^a q^n}{1-q^n} \; .
\end{equation}

With the help of $\eta(\tau)$ and $G^{}_4(\tau)$, one can define a modular-invariant function which is called the Klein $j$-function as
\begin{eqnarray}
    j(\tau) = \frac{3^6 5^3}{\pi^{12}} \frac{G_4(\tau)^3}{\eta(\tau)^{24}} \; .
    \label{eq:Klein}
\end{eqnarray}

\section{Series expansion of the scalar potential} \label{app:derivation}

In order to derive the series expansion of the scalar potential in the vicinity of $x = 0$, we should first calculate the derivatives of $G_2(\tau)$ and $\eta(\tau)$ in terms of $\tau$. Ramanujan identities indicate that the derivatives of Eisenstein series can be expressed as the linear combinations of the first few Eisenstein series. Adopting an alternative notation $E_{2k} \equiv G_{2k}/[2\zeta(2k)]$, it is found that
\begin{equation}
\begin{aligned}
 \frac{d E^{}_2}{d \tau} & =2\pi{\rm i}\frac{E^2_2-E^{}_4}{12} \; ,\\
 \frac{d E^{}_4}{d \tau} & =2\pi{\rm i}\frac{E^{}_2 E^{}_4-E^{}_6}{3} \; ,\\
 \frac{d E^{}_6}{d \tau} & =2\pi{\rm i}\frac{E^{}_2 E^{}_6-E^2_4}{2} \; .
\end{aligned}
\label{eq:rama-inden}
\end{equation}
With the help of Eq.~(\ref{eq:rama-inden}), we can obtain derivatives of any order of Eisenstein series with respect to $\tau$. Here we pay particular attention to $G_2$. At the fixed point $\tau = \I$, we have 
\begin{equation}
\begin{split}
    G_2|_{\tau = \I} & = \pi \; , \\
    \left.\frac{{\rm d}G_2}{{\rm d}\tau} \right|_{\tau = \I} & = 
    \frac{ \pi \I}{6}\left[3 - \frac{\Gamma^8(1/4)}{64\pi^4}\right] \; , \\
    \left.\frac{{\rm d}^2G_2}{{\rm d}\tau^2} \right|_{\tau = \I} & = 
    -\frac{\pi}{2}\left[1 - \frac{\Gamma^8(1/4)}{64\pi^4}\right] \; , \\
    \left.\frac{{\rm d}^3G_2}{{\rm d}\tau^3} \right|_{\tau = \I} & = 
    -\frac{\pi \I}{4}\left[3 - \frac{3\Gamma^8(1/4)}{32\pi^4} - \frac{\Gamma^{16}(1/4)}{4096\pi^8}\right] \; , \\
    \left.\frac{{\rm d}^4G_2}{{\rm d}\tau^4} \right|_{\tau = \I} & = 
    \frac{\pi}{2}\left[3 - \frac{5\Gamma^8(1/4)}{32\pi^4} - \frac{5\Gamma^{16}(1/4)}{4096\pi^8}\right] \; ,
    \end{split}
    \label{eq:DG2}
\end{equation}
where $\Gamma(z) \equiv \int^\infty_0 t^{z-1} e^{-t} {\rm d}t$ denotes the Euler $\Gamma$-function. $G_2$ can be related to $\eta(\tau)$ by
\begin{equation}
    \frac{\eta^\prime(\tau)}{\eta(\tau)} = \frac{\I}{4\pi}G_2(\tau) \; .
\end{equation}
Given that $\eta(\I) = \Gamma(1/4)/(2\pi^{3/4})$, we can also evaluate the derivatives of $\eta(\tau)$ at $\tau = \I$ as
\begin{equation}
    \begin{split}
        \left.\frac{{\rm d}\eta}{{\rm d}\tau} \right|_{\tau = \I} & = \frac{\I \Gamma(1/4)}{8\pi^{3/4}} \; \\
        \left.\frac{{\rm d}^2\eta}{{\rm d}\tau^2} \right|_{\tau = \I} & = -\frac{\Gamma(1/4)[288\pi^4 - \Gamma^8(1/4)]}{3072\pi^{19/4}} \; , \\
        \left.\frac{{\rm d}^3\eta}{{\rm d}\tau^3} \right|_{\tau = \I} & = -\frac{5\I\Gamma(1/4)[96\pi^4 - \Gamma^8(1/4)]}{4096\pi^{19/4}} \; , \\
        \left.\frac{{\rm d}^4\eta}{{\rm d}\tau^4} \right|_{\tau = \I} & = \frac{\Gamma(1/4)[322560\pi^8 - 6720\pi^4\Gamma^8(1/4)-11\Gamma^{16}(1/4)]}{1572864\pi^{35/4}} \; .
    \end{split}
    \label{eq:Deta}
\end{equation}
Using Eqs.~(\ref{eq:DG2}) and (\ref{eq:Deta}), and keeping in mind that $\tau = {\rm exp} [2\I\,{\rm arctan}(e^{x/\sqrt{3}})]$ is satisfied on the lower boundary of the fundamental domain, we can derive the Taylor expansion of the scalar potential $V(x)$ at $x = 0$ up to the fourth order of $x$, which is just Eq.~\eqref{eq:scalar-expand} in the main text.

\section{Evading the cosmological constant problem by introducing a waterfall field} \label{app:waterfall}

In this appendix, we provide a mechanism to eliminate the large vacuum energy at $\tau = \omega$ after inflation by introducing an additional ``waterfall'' field direction~\cite{Antoniadis:2020stf}. This direction  may originate from another modulus field in a multi-moduli scenario. We consider a toy model by introducing a real scalar field $\sigma \equiv y\Lambda_V$, which leads to the contribution
\begin{equation}
    \frac{\Delta V(x,y)}{\Lambda_V^4} = \frac{1}{2}[-\widehat{\mu}^2_{} +h(x)]y^2+\frac{\lambda}{4}y^4 \; ,
    \label{eq:waterfallp}
\end{equation}
where $h(x)$ is a dimensionless function of the dimensionless inflaton field $x$ (analogous to the field $x$ introduced in this the paper), and $\widehat{\mu}$ and $\lambda$ are dimensionless coefficients. The effective mass squared of $\sigma$ is then $m_\sigma^2 = \Lambda^2_{V}[-\widehat{\mu}^2_{} +h(x)]$. Then the entire potential becomes $V_{\rm tot}(x,y) = V(x) + \Delta V(x,y)$. It is worth mentioning that the modular invariance of the scalar potential should be preserved when additional moduli are included. Therefore, Eq.~(\ref{eq:waterfallp}) should be regarded as the first few terms of the series expansion of the potential.

\begin{figure}[!t]
	\centering
\includegraphics[width=0.9\linewidth]{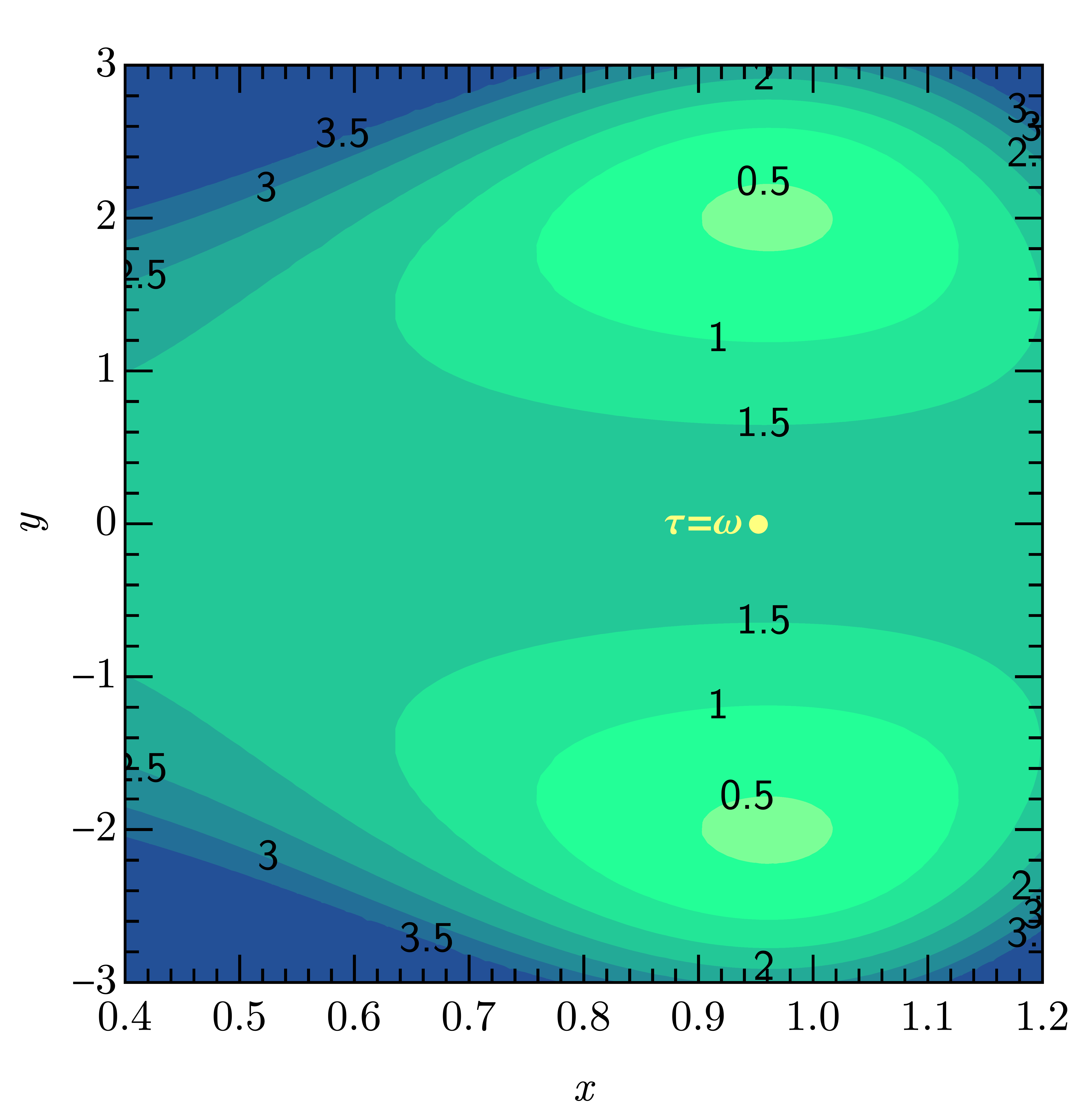}
	\caption{Contour plot of the two field toy model potential $V_{\rm tot} = V(x) + \Delta V(x,y)$ in the $x-y$ plane. $V(x)$ is obtained by substituting $\tau = {\rm exp} [2\I\,{\rm arctan}(e^{x/\sqrt{3}})]$ into Eq.~(\ref{eq:single-ponten-2}) and choosing $\mathfa=0.04$ and $\alpha = -2.73 \times 10^{-3}$. For $\Delta V(x,y)$, we assume $\widehat{\mu}^2 = 1.3$, $\lambda=0.32$, and $h(x)=2.4(x^2-0.925)^2$. Numbers on the contours correspond to the values of $V_{\rm tot}$ in units of $\Lambda_V^4$. The yellow dot denotes $\tau = \omega$.}
	\label{fig:waterfall}
\end{figure}

We can choose an appropriate form of $h(x)$, so that $m^2_\sigma > 0 $ during the inflation phase. Then $y$ is stabilised at $y = 0$ as the inflaton rolls down. At the end of inflation, $m^2_\sigma$ becomes negative, inducing an instability in the $y$-direction. Consequently, the waterfall field $y$ will rapidly fall along this direction and the complete potential will reach a minimum with lower energy. As an illustrative example, we assume $\widehat{\mu}^2 = 1.3$, $\lambda=0.32$, and $h(x)=2.4(x^2-0.925)^2$. The shape of $V_{\rm tot}$ is shown in \figref{fig:waterfall}, where we can observe that $\tau = \omega$ is indeed no longer a stable minimum. Instead, new vacua with lower energy appear in the $y$-direction. In such a scenario one can evade the cosmological constant problem by introducing a waterfall field.

\newpage
\bibliographystyle{JHEP}
\bibliography{Ref}
\end{document}